\documentclass{vldb}

\usepackage{booktabs} % For formal tables
\usepackage{appendix}
\usepackage{graphicx}
\usepackage{balance} 
\usepackage{color}
\usepackage[utf8]{inputenc}
\usepackage{listings} 
\usepackage{comment}
\usepackage{mdwlist}
\usepackage{tabularx} 
\usepackage{tikz}
\usepackage{xspace}
\usepackage{cleveref}
\usepackage[binary-units]{siunitx}
\usepackage{paralist}
\usepackage{todonotes}
\usepackage{enumitem}
\usepackage{pgfplots}
\usepackage{subfig}
\usepackage{hyperref}
\usepackage{verbatim}

\vldbTitle{DBPal: Weak Supervision for Learning a Natural Language Interface to Databases}
\vldbAuthors{Nathaniel Weir, Andrew Crotty, Alex Galakatos, Amir Ilkhechi, Shekar Ramaswamy, Rohin Bhushan, Ugur Cetintemel, Prasetya Utama, Nadja Geisler, Benjamin H\"attasch, Steffen Eger, Carsten Binnig}
\vldbYear{2019}

\newcommand{\system}{{\sc DBPal}}

\begin{document}
\title{DBPal: Weak Supervision for Learning a \\Natural Language Interface to Databases}

\numberofauthors{2} 

\author{
\alignauthor Nathaniel Weir, Andrew Crotty,\\Alex Galakatos, Amir Ilkhechi,\\Shekar Ramaswamy, Rohin Bhushan,\\Ugur Cetintemel\\
\affaddr{Brown University, USA}
\alignauthor Prasetya Utama, Nadja Geisler,\\Benjamin H\"attasch, Steffen Eger,\\Carsten Binnig\\
\affaddr{TU Darmstadt, Germany}
}

\maketitle

\begin{abstract}
This paper describes \system{}, a new system to translate natural language utterances into SQL statements using a neural machine translation model. While other recent approaches use neural machine translation to implement a Natural Language Interface to Databases (NLIDB), existing techniques rely on supervised learning with manually curated training data, which results in substantial overhead for supporting each new database schema. In order to avoid this issue, \system{} implements a novel training pipeline based on weak supervision that synthesizes all training data from a given database schema. In our evaluation, we show that \system{} can outperform existing rule-based NLIDBs while achieving comparable performance to other NLIDBs that leverage deep neural network models without relying on manually curated training data for every new database schema.
\end{abstract}

\section{Introduction}
\label{sec:intro}
\textbf{Motivation:}
Structured query language (SQL), despite its expressiveness, may hinder users with little or no relational database knowledge from exploring and making use of the data stored in a DBMS.
In order to effectively analyze such data, users are required to have prior knowledge of the syntax and semantics of SQL.
These requirements set ``a high bar for entry'' for democratized data exploration and have therefore triggered new efforts to develop alternative interfaces that allow non-technical users to explore and interact with their data more conveniently.
Despite the recent popularity of visual data exploration tools (e.g., Tableau~\cite{tableau}, Vizdom~\cite{vizdom}), Natural Language Interfaces to Databases (NLIDBs) have emerged as a highly promising alternative, since they enable users to pose questions in a concise and intuitive manner.

\textbf{Contributions:}
Understanding natural language (NL) questions and translating them accurately into SQL is a complicated task, and thus NLIDBs have not yet made their way into mainstream commercial products.
This paper describes \system{}, a natural language interface to databases with improved robustness to linguistic variations.
In order to provide a more robust NL interface, \system{} leverages a deep neural network model, which has become a standard for machine translation tasks. 
While many other recent efforts also use deep models to implement NLIDBs~\cite{DBLP:conf/acl/IyerKCKZ17,Wang2015BuildingAS,DBLP:journals/corr/abs-1711-04436}, they commonly rely on supervised learning approaches that require substantial amounts of training data, particularly for sophisticated neural architectures such as deep sequence-to-sequence models. 

The aforementioned approaches have largely ignored this problem and assumed the availability of manually curated training sets (e.g., via crowdsourcing). 
As such, additional manual effort is needed for each new database schema, which severely limits the portability of these approaches to new domains. 
In order to address this fundamental limitation, we have built \system{} as a complete system that enables users to build robust NL interfaces for different databases with low manual overhead. 

At its core, \system{} implements a novel training pipeline for NLIDBs that synthesizes its training data using the principle of \textit{weak supervision}~\cite{DBLP:journals/ai/CravenDFMMNS00,DBLP:conf/sigir/DehghaniZSKC17}.
The basic idea of weak supervision is to leverage various heuristics and existing datasets to automatically generate large (and potentially noisy) training datasets instead of handcrafting them.

In its basic form, only the database schema is required as input in order to generate a large collection of pairs of natural language queries and their corresponding SQL statements that can then be used to train our language translation model. 
In order to maximize our coverage across natural language variations, we use additional input sources to automatically augment the training data using a collection of techniques.
One such augmentation step, for example, is an automatic paraphrasing process using an off-the-shelf paraphrasing database~\cite{DBLP:conf/acl/PavlickC16a}. 

\textbf{Outline:}
The remainder of this paper is organized as follows.
In Section \ref{sec:overview}, we first introduce the overall system architecture of \system{}.
Afterwards, in Section \ref{sec:training}, we describe the details of the novel training pipeline of \system{} based on weak supervision.
In order to show the accuracy of \system{} as well as its robustness, we present initial results of our evaluation in Section \ref{sec:eval}.
Finally, we describe promising future directions in Section~\ref{sec:future}.
\section{Overview}
\label{sec:overview}
\begin{figure}
\centering
\includegraphics[width=0.45\textwidth]{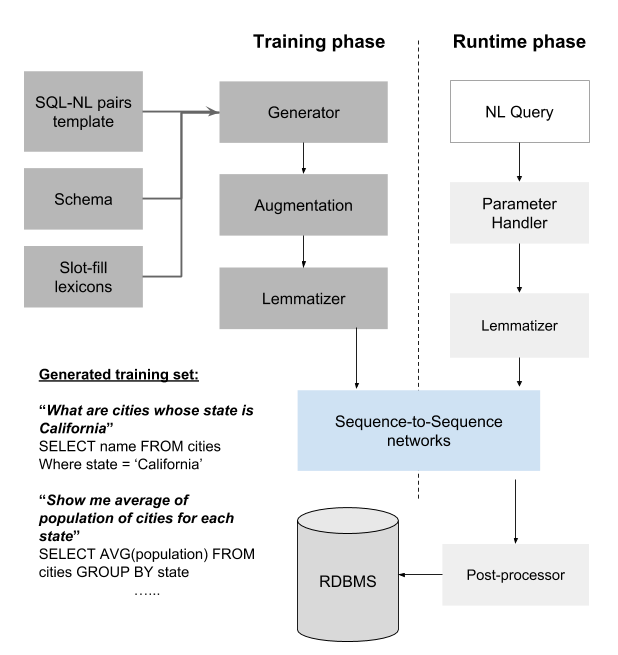}
%\vspace{-2.5ex}
\caption{\system{}'s Training and Runtime Phases
\vspace{-3.5ex}
}
\label{fig:translator}
\end{figure}

Figure \ref{fig:translator} shows an overview of the architecture of \system{}.
At the core of \system{} is a neural machine translation model (i.e., a sequence-to-sequence model).

The most important aspect of \system{} is the novel training pipeline based on weak supervision that automatically generates training data used for training the translation model.
The basic flow of the training pipeline is shown on the left-hand side of Figure~\ref{fig:translator}.
In the following, we describe the training pipeline and focus in particular on the data generation framework. 
The details of the full training pipeline will be explained in Section~\ref{sec:training}.

In the first step, the \textit{Generator} uses the database schema along with a set of seed templates that describe typical NL-SQL pairs to generate an initial training set of NL-SQL pairs. 
In the second step, \textit{Augmentation}, the training data generation pipeline then automatically adds to the initial training set of NL-SQL pairs by leveraging existing general-purpose data sources and models to linguistically modify the NL part of each pair.

Furthermore, the runtime phase is comprised of multiple components, as shown on the right-hand side of Figure \ref{fig:translator}.
The \textit{Parameter Handler} is responsible for replacing the constants in the input NL query with placeholders to make the translation model independent from the actual database content and avoid needing to retrain the model if the database is updated.
For example, for the input query shown in Figure \ref{fig:translator} (i.e., \textit{``What are cities whose state is California?''}), the \textit{Parameter Handler} replaces \textit{``California''} with the appropriate schema element using the placeholder \texttt{@STATE}.
The neural model then works on these anonymized NL input queries and creates output SQL queries, which also contain placeholders.
In the example in Figure~\ref{fig:translator}, the output of the neural translator is: \texttt{SELECT name FROM cities WHERE state = @STATE}.
The task of the \textit{Post-processor} is then to replace the placeholders again with the actual constants such that the SQL query can be executed against the database.

\section{Training Pipeline}
\label{sec:training}
In the following, we describe the training data generation procedure implemented in \system{}, which is at the core of our weakly supervised training pipeline.
In this pipeline, an initial instantiation step first generates a simple set of NL-SQL pairs for a given schema.
Afterwards, an augmentation step that is based on existing language models (e.g., for automatic paraphrasing) generates further linguistic variants for each NL-SQL pair to cover a wider range of variations for possible natural language questions.

\subsection{Data Instantiation}
The main observation of the instantiation step is that SQL, as opposed to natural language, has significantly less expressivity.
We therefore use query templates to instantiate different possible SQL queries that a user might phrase against a given database schema, such as:\\

\vspace{-1ex}
\noindent \textit{Select \textsf{\{Attribute\}(s)} From  \textsf{\{Table\}} Where \textsf{\{Filter\}}}\\
\vspace{-1ex}

The SQL templates cover a variety of query types, from simple \texttt{SELECT-FROM-WHERE} queries to more complex group-by aggregation queries as well as some simple nested queries.
For each SQL template, we define one or more NL templates as counterparts for direct translation, such as:\\

\vspace{-1ex}
\noindent \textit{\{SelectPhrase\} \textsf{\{Attribute\}(s)} \{FromPhrase\} \textsf{\{Table\}(s)} \\
\{WherePhrase\} \textsf{\{Filter\}}}\\
\vspace{-1ex}

To take into account the expressivity of NL compared to SQL, our templates contain slots for speech variation (e.g., \textit{SelectPhrase}, \textit{FromPhrase}, \textit{WherePhrase}) in addition to slots for database objects (e.g., tables, attributes). %These slots can be instantiated using our slot-fill lexicons.
Then, to instantiate the initial training set, the generator repeatedly instantiates each of our natural language templates by filling in their slots. 
Table, column, and filter slots are filled using information from the database's schema, while a diverse array of natural language slots are filled using manually crafted dictionaries of synonymous words and phrases. 
%\fb{The corpus of possible training items created in this way already covers a wide spectrum of possible query formulations that a user might give.} 
For example, the phrases \textit{``what is''} or \textit{``show me''} can be used to instantiate the \textit{SelectPhrase}.
A fully instantiated NL-SQL pair might look like:\\

\vspace{-1ex}
\noindent \textbf{NL:} \textit{Show the names of all patients with age 20.}\\
\textbf{SQL:} \texttt{SELECT name FROM patients WHERE age = 20}\\
\vspace{-1ex}

An important part of training data instantiation is balancing the number of NL-SQL pairs that are instantiated per template.
%from the filling in of templates with different numbers of slots, given the addition of a new slot exponentiates the number of instances that can be generated.
If we naively replace the slots of a query template with all possible combinations of slot instances (e.g., all attribute combinations of the schema), then instances that result from templates with more slots would dominate the training set and bias the translation model.
An imbalance of instances can result in a biased training set where the model would prefer certain translations over others only due to the fact that certain variations appear more often.
%We therefore randomly sample from the possible instances to get a good coverage of different queries and to keep the number of instances per query template balanced.
%replaced post to pre

Finally, in the current prototype, for each initial NL template, we additionally provide some manually curated paraphrased NL templates that follow particular paraphrasing techniques~\cite{DBLP:journals/pdln/VilaMR11}, covering categories such as syntactical, lexical, and morphological paraphrasing. 
Importantly, the paraphrased templates can be applied to instantiate the training data for any given schema, and the instantiated NL-SQL pairs are also automatically paraphrased during automatic data augmentation.

\subsection{Data Augmentation}
In order to make the query translation model more robust to linguistic variation (i.e., how a user might phrase an input NL query), we apply the following augmentation steps for each instantiated NL-SQL pair.

First, we augment the training set by generating duplicate NL-SQL pairs.
This process involves randomly selecting words/subphrases of the NL query and paraphrasing them using the Paraphrase Database (PPDB) \cite{DBLP:conf/acl/PavlickC16a} as a lexical resource, for example:\\

\makebox[0.48\textwidth][c]{
\centering
\begin{minipage}[c]{0.48\textwidth}
\noindent \textbf{Input NL Query:} \\
\textit{\underline{Show} the names of all patients with age @AGE.}
\vspace{1.5ex}\\
\noindent \textbf{PPDB Output:} \\
demonstrate, showcase, display, indicate, lay
\vspace{1.5ex}\\
\noindent \textbf{Paraphrased NL Query:} \\
\textit{\underline{Display} the names of all patients with age @AGE.}\\
\end{minipage}}

PPDB is an automatically extracted database containing millions of paraphrases in $27$ different languages.
\system{} uses PPDB's English corpus, which provides over $220$ million paraphrase pairs consisting of $73$ million phrasal and $8$ million lexical paraphrases, as well as $140$ million paraphrase patterns, which capture a wide range of meaning-preserving syntactic transformations.
The paraphrases are extracted from bilingual parallel corpora totaling over $100$ million sentence pairs and over $2$ billion English words. 

During paraphrasing, we randomly replace words/subphrases of the input NL query with available paraphrases provided by PPDB. 
For example, searching in PPDB for a paraphrase of the word \textit{enumerate}, as in \textit{``enumerate the names of patients with age 80''}, we get suggestions such as \textit{``list''} or \textit{``identify''} as alternatives.

Second, to make the translation more robust to missing or implicit context, we randomly drop words/subphrases from the NL training queries.
For example, a user might ask for \textit{``patients with flu''} instead of \textit{``patients diagnosed with flu''}, where the referenced attribute is never explicitly stated.
Similar to paraphrasing, an interesting question is: which words/subphrases should be removed and how frequently to remove them?
Again, aggressively removing words from query copies increases the training data size, since more variations are generated.
On the other hand, however, we again might introduce noisy training data that leads to a drop in translation accuracy.

In order to tune how aggressively we apply removal as well as the other augmentation techniques, we provide parameters for the data generation process.
Tuning such parameters is similar to hyper-parameter tuning in machine learning, and we therefore plan to explore routes in the future that investigate how to automatically tune the data generation parameters.
\begin{table*}[]
\centering
\small
\begin{tabular}{|l|r|r|}
\hline
& \textbf{Patients} & \textbf{GeoQuery} \\ \hline
\begin{tabular}[c]{@{}l@{}}\textbf{NaLIR} \textbf{(no feedback)}\end{tabular} & 15.60\% & 7.14\%\\ \hline
\begin{tabular}[c]{@{}l@{}}\textbf{NaLIR} \textbf{(feedback)}\end{tabular} & 21.42\% & N/A \\ \hline
\textbf{NSP} &N/A &\textbf{83.9}\%        \\ \hline
\textbf{\system{} (no augmentation)} & 74.80\% & 38.60\%          \\ \hline
\textbf{\system{} (full pipeline)} & \textbf{75.93}\% & 55.40\%          \\ \hline
\end{tabular}
\vspace{1mm}
\caption{Accuracy comparison between \system{} and other baselines on both benchmarks.}
%\vspace{-3.5ex}
\label{table:results}
\end{table*}

\begin{table*}[]
\centering
\small
\begin{tabular}{|l|r|r|r|r|r|r|r|}
\hline
 & \textbf{Naive} & \textbf{Syntactic} & \textbf{Lexical} & \textbf{Morphological} & \textbf{Semantic} & \textbf{Missing} & \textbf{Mixed}\\ \hline
\begin{tabular}[c]{@{}l@{}}\textbf{NaLIR} \textbf{(no feedback)}\end{tabular} & 19.29\% & 28.07\% & 14.03\% & 17.54\% & 7.01\% & 5.77\% & 17.54\%\\ \hline
\begin{tabular}[c]{@{}l@{}}\textbf{NaLIR} \textbf{(feedback)}\end{tabular} & 21.05\% & 38.59\% & 14.03\% & 19.29\% & 7.01\% & 5.77\% & 22.80\%\\ \hline
\textbf{\system{} (full pipeline)} & \textbf{96.49}\% & \textbf{94.7}\% & \textbf{75.43}\% & \textbf{85.96}\% & \textbf{57.89}\% & \textbf{36.84}\% & \textbf{84.20}\%\\ \hline
\end{tabular}
\vspace{1mm}
\caption{Accuracy breakdown between \system{} and other baselines for the Patients benchmark.}
\vspace{-3.0ex}
\label{table:breakdown}
\end{table*}

\section{Initial Results}
\label{sec:eval}
In the following, we present our initial experimental results with \system{}.
In all experiments, we compare against a neural semantic parser (NSP)~\cite{DBLP:conf/acl/IyerKCKZ17} approach, which also leverages a deep model for the translation process.
However, unlike \system{}, NSP requires a manually curated training set for each new database schema.
As a second baseline, we also used NaLIR~\cite{nalir1,nalir2}, a state-of-the-art rule-based NLIDB that requires no training data to support a new schema.

The first part of our evaluation uses the well-known GeoQuery benchmark, but this benchmark does not explicitly test different linguistic variations to measure robustness.
For testing different linguistic variants, we curated a new benchmark, called the Patients~\footnote{\url{https://datamanagementlab.github.io/ParaphraseBench/}} benchmark, that covers different linguistic variations for the user NL input and maps it to an expected SQL output.

\subsection{Exp. 1: Overall Results}
We evaluated the performance of all NLIDB systems in terms of their accuracy, defined as the number of natural language queries translated correctly over the total number of queries in the test set.
Correctness is determined by whether the yielded records from a query's execution in the DBMS contain the information that is requested by the query intent.
The correctness criteria is relaxed by also considering execution results that consist of supersets of the requested columns to be correct.
We argue that, in practice, users are still able to retrieve the desired information by examining all columns of returned rows.
Table~\ref{table:results} summarizes the accuracy measures of all NLIDB systems on the two benchmark datasets using this correctness criterion.

First, we note that \system{} outperforms NaLIR, which relies heavily on user feedback when it cannot automatically find a valid SQL translation.
We ran NaLIR in the baseline non-interactive mode as well as interactive mode, where we provide perfect feedback (i.e., users always make the correct choices if NaLIR asks for feedback).
With interactive feedback, we observe a slight increase in accuracy on the Patients benchmark, but we did not run NaLIR in interactive mode on GeoQuery due to the high manual effort of providing feedback for more than $300$ queries.

Second, we observe that NSP, which requires the costly curation of a training set of NL-SQL pairs, unsurprisingly achieves the highest accuracy on its given domain of GeoQuery.
We could not evaluate NSP on the Patients benchmark, since no manually curated training data was available.
It is interesting to note that NSP performs well on GeoQuery while \system{} only achieves approximately $50\%$ accuracy.
However, after analyzing the manual training data of NSP, we note that it contains many of the structurally similar NL-SQL pairs in their training set that are also in the test set of GeoQuery.
Furthermore, upon anonymization of database values, it becomes clear that identical queries appear in both the training and testing sets.
Thus, the learned model is heavily overfit to the training patterns.

Another observation is that \system{} has a lower accuracy on GeoQuery than on Patients. 
This is due to the fact that GeoQuery contains a huge fraction of rather complex joins and nested queries that are currently not supported well in \system{}.
When removing those queries from GeoQuery, \system{} actually performs comparably to NSP.
%Another complexity in GeoQuery is that it also contains domain-specific natural language utterances that are not easily obtainable via existing automatic paraphrasing methods. For example, the query 'what rivers \emph{run through} California' is not easily translatable by \system{} because the term `run through' is very difficult to identify with its proper semantic role. 

Finally, we analyzed the benefit of the data augmentation step of our training data generation, which produces more NL-SQL pairs with a large amount of variety.
For the Patients benchmark, we can see that this contributes only minor improvements to the performance, while for the GeoQuery dataset, the accuracy is almost $20\%$ higher.
This is due to the higher complexity of this database schema, which consists of multiple tables with relationships that bring about richer semantics when asking questions. 

\subsection{Exp. 2: Performance Breakdown}
For the Patients benchmark, we show the breakdown of different NL variants in Table~\ref{table:breakdown}.
We see that NaLIR achieves a similar accuracy of $11/57=19.29\%$ without user feedback and a slightly higher accuracy of $12/57=21.05\%$ with user feedback for the naive testing set. 
Furthermore, in most testing sets, perfect user feedback does not significantly improve the accuracy of NaLIR, since NaLIR relies on an off-the-shelf dependency parser library that performs reasonably well only on well-structured sentences.
Therefore, most of NaLIR's failure cases are due to dependency parsing errors caused by ill-formed, incomplete, or keyword-based queries.
Moreover, NaLIR often fails to find correct candidate mappings of query tokens to schema elements due to paraphrased inputs.
Both of these problems cannot be repaired by user feedback, given the translation procedure fails before any feedback can be provided.
\section{Future Work}
\label{sec:future}
In the immediate future, we plan to explore extensions to the training data instantiation and augmentation processes by creating additional templates and lexicons to cover more linguistic variations, as well as to better support even more complex SQL queries (e.g., joins, nested queries).
Additionally, \system{} currently has no good mechanisms for explaining results to the user, suggesting possible corrections if the translation was incorrect, or incorporating user feedback to make corrections.
We therefore plan to explore ways to enable users to incrementally build and refine queries in a conversational chatbot-like interface, where the system can ask for clarifications if the model cannot translate a given input query directly.
This feature will also help to improve the overall user experience by leveraging contextual clues from a series of interactions without requiring the user to explicitly reiterate elements from past queries.

Longer term, we believe that an exciting opportunity exists to expand \system{} into a full-fledged data science platform that will allow domain experts to interactively explore large datasets using only natural language~\cite{ava}.
Again, in contrast to the typical notion of one-shot SQL queries currently taken by \system{}, data science is an iterative, session-driven process where a user repeatedly modifies a query or machine learning model after examining intermediate results until finally arriving at some desired insight, which will therefore necessitate a more conversational interface.
Such a system would require the development of new techniques for providing progressive results~\cite{progressive1,progressive2} by extending past work on traditional SQL-style queries~\cite{idea-hilda,idea-vldb} and machine learning models~\cite{alpinemeadow}.

Finally, we believe there are also interesting opportunities related to different data models (e.g., time series~\cite{timeseries}) and new user interfaces (e.g., query-by-voice~\cite{echoquery}).

\begin{scriptsize}
\bibliographystyle{abbrv}
\bibliography{bib}  
\end{scriptsize}

\end{document}